\documentstyle[12pt,aaspp4,flushrt]{article}
\makeatletter

\newlength{\bibhang}
\let\@internalcite\cite
\def\cite{\let\@citeleft(\let\@citeright)%
    \@ifstar{\citeyear}{\citefull}}
\def\citenp{\let\@citeleft\relax\let\@citeright\relax
    \@ifstar{\citeyear}{\citefull}}
\def\citefull{\def\astroncite##1##2{##1~##2}\@internalcite}
\def\citeyear{\def\astroncite##1##2{##2}\@internalcite}

\def\@citex[#1]#2{\if@filesw\immediate\write\@auxout{\string\citation{#2}}\fi
  \def\@citea{}\@cite{\@for\@citeb:=#2\do
    {\@citea\def\@citea{; }\@ifundefined
       {b@\@citeb}{{\bf ?}\@warning
       {Citation `\@citeb' on page \thepage \space undefined}}%
{\csname b@\@citeb\endcsname}}}{#1}}

\def\@cite#1#2{\@citeleft#1\if@tempswa , #2\fi\@citeright}
\def\@biblabel#1{}

\makeatother

\newcommand{\msun}{$M_\odot$} 
\newcommand{\persec}{\mbox{$\second^{-1}$}}
\newcommand{\percm}{\mbox{$\cm^{-2}$}}
\newcommand{\ppm}{\mbox{$\pm$}}
\newcommand{\cgsflux}{\erg\percm\persec}
\newcommand{\cgslum}{\erg\persec}

\newcommand{\approxlt}{\mbox{$\lesssim$}}

\def\etal{{et~al.}}
\def\eg{{e.g.}}
\def\ie{{i.e.}}
\def\x1608{{4U~1608$-$522}}
\def\cenx4{{Cen~X$-$4}}
\newcommand{\nh}{\mbox{$N_{\rm H}$}}
\newcommand{\nhtt}{\mbox{$N_{\rm H, 22}$}}
\def\aql{{Aql~X$-$1}}
\def\chisqr{\mbox{$\chi^2$}}
\newcommand{\ud}[2]{\mbox{$^{+ #1}_{- #2}$}}
\newcommand{\ee}[1]{\mbox{$10^{#1}$}}
\newcommand{\tee}[1]{\mbox{$\times 10^{#1}$}}
\newcommand{\tento}[1]{\mbox{$10^{#1}$}}
\newcommand{\keV}{\mbox{$\rm\,keV$}}

\newcommand{\cm}{\mbox{$\rm\,cm$}}
\newcommand{\km}{\mbox{$\rm\,km$}}
\newcommand{\second}{\mbox{$\rm\,s$}}

\newcommand{\erg}{\mbox{$\rm\,erg$}}
\newcommand{\gauss}{\mbox{$\rm\,G$}}

\newcommand{\axaf}{{\em AXAF\/}}
\newcommand{\rosat}{{\em ROSAT\/}}
\newcommand{\asca}{{\em ASCA\/}}

\newcommand{\xmm}{{\em XMM\/}}
\newcommand{\beppo}{{\em BeppoSAX\/}}
\newcommand{\kpc}{\mbox{$\rm\,kpc$}}

\newcommand{\Teff}{T_{\rm eff}}
\received{ApJ, Sept 22, 1998}
\accepted{Oct 19,  1998}
\begin{document}

\title{The Thermal X-ray Spectra of \cenx4, \aql , and \x1608\ in
Quiescence}
\author{Robert E. Rutledge\altaffilmark{1}, 
Lars Bildsten\altaffilmark{2}, Edward F. Brown\altaffilmark{2}, 
George G. Pavlov\altaffilmark{3}, 
\\Vyatcheslav  E. Zavlin\altaffilmark{4}}
\altaffiltext{1}{
Space Radiation Laboratory, California Institute of Technology, MS 220-47, Pasadena, CA 91125;
rutledge@srl.caltech.edu}
\altaffiltext{2}{
Department of Physics and Department of Astronomy, 601 Campbell Hall,
Mail Code 3411, University of California at Berkeley, Berkeley, CA
94720; bildsten@fire.berkeley.edu, ebrown@astron.berkeley.edu}
\altaffiltext{3}{
The Pennsylvania State University, 525 Davey Lab, University Park, PA
16802; pavlov@astro.psu.edu}
\altaffiltext{4}{
Max-Planck-Institut f\"ur Extraterrestrische Physik, D-85740 Garching,
Germany; zavlin@xray.mpe.mpg.de}

\begin{abstract}

We re-analyze the available X-ray spectral data of the type~I bursting
neutron star transients \aql, \cenx4, and \x1608\ using realistic
hydrogen atmosphere models.  Previous spectral fits assumed a
blackbody spectrum; because the free-free dominated photospheric
opacity decreases with increasing frequency, blackbody spectral fits
overestimate the effective temperature and underestimate, by as much
as two orders of magnitude, the emitting area.  Hydrogen atmosphere
spectral models, when fit to the available observational data, imply
systematically larger emission area radii, consistent with the
canonical 10~km radius of a neutron star.  This suggests that a
substantial fraction of the quiescent luminosity is thermal emission
from the surface of the neutron star.  The magnitude of the equivalent
hydrogen column density toward these systems, however, presents a
considerable systematic uncertainty, which can only be eliminated by
high signal-to-noise X-ray spectral measurements (\eg , with \axaf\ or
\xmm ) which would permit simultaneous determination of the equivalent
hydrogen column density, emission area, and thermal temperature.

\end{abstract}


\section{Introduction} \label{sec:intro}

What distinguishes a transient X-ray binary from a persistent one is
ill-defined physically, but observationally it may be defined as an
object whose flux changes by more than 2--3 orders of magnitude (for
recent reviews, see \citenp{tanakalewin95,chen97,campana98b}).  Many
neutron star (NS) and black hole (BH) transients go through X-ray
outbursts, separated by long periods (months to years) of relative
quiescence.  The origin of these outbursts remains under debate,
although most mechanisms rely in some form on an accretion instability
\cite{jvp96,king96}.

Whatever the cause of an X-ray outburst, these objects eventually
return to quiescence.  Several NS and BH transients have been detected
in quiescence, with typical luminosities for the NSs being
$\sim\tento{33}\erg\second^{-1}$.  Here, we consider the low-magnetic
field (as inferred from the presence of type~I X-ray bursts) NS
transients.

The first NS transient detected in quiescence was \cenx4\
\cite[hereafter VP87]{jvp87}.  It was argued that radiative cooling
from the NS surface could not be the emission mechanism, based on two
observations.  First, the inferred radius of the emitting area, using
a blackbody (BB) spectrum, assuming a distance of d=2.3 \kpc
\cite{blair84}, was $0.83\ud{0.72}{0.40}\km$ (90\% confidence),
smaller than that expected from a $\sim$10~km NS. Second, the
quiescent luminosity observed by the {\em Einstein\/} IPC was a factor
of $\sim 2\mbox{--}4$ lower than the later {\em EXOSAT\/} CMA
observation (the exact factor depends on the assumed spectrum);
because the core temperature cannot appreciably change in just a few
years, the thermal heat flux from a hot core should also remain stable
over this time scale.  It was therefore suggested that the emission is
caused by continued accretion over a fraction of the surface.

More recently, quiescent X-ray spectral measurements have been made of
\aql\ with the \rosat/PSPC \cite{verbunt94} and of \cenx4\ and \x1608\
with \asca\ \cite{asai96b}.  The X-ray spectrum of \aql\ (0.4--2.4\keV)
was consistent with a BB spectrum, a bremsstrahlung spectrum, or a
pure power-law spectrum \cite{verbunt94}.  For \x1608, the spectrum
(0.5--10.0\keV) was consistent with a BB ($kT_{\rm BB}\sim
0.2\mbox{--}0.3$ keV), a thermal Raymond-Smith model ($kT =
0.32\ud{0.18}{0.5}$ keV), or a very steep power-law (photon index
$6\ud{1}{2}$).  Similar observations of \cenx4\ with \asca\ found its
X-ray spectrum consistent with these same models, but with an
additional power-law component (photon index $\sim2.0$) above 5.0\keV\
(recent observations with \beppo\ of \aql\ in quiescence also revealed
a power-law tail; \citenp{campana98a}).  In all three sources, BB fits
implied an emission area of radius $\sim 1\km$.  A thermally emitting
region this small is difficult to explain, unless one assumes that the
NS accretes onto only a fraction of its surface during quiescence.

Aside from luminosity due to low-level accretion, thermal emission
from transiently accreting NSs would be observable during quiescence
if the NS core were sufficiently hot (VP87;
\citenp{verbunt94,asai96a,campana98b}).  H/He burning alone cannot
heat the core to the interior temperatures of a steadily accreting
star.  The heat released from hydrogen/helium burning in the upper
atmosphere leaves immediately during the unstable burning
\cite{hanawa86,fujimoto87}, and the time between accretion outbursts
is much longer than the cooling time of the NS atmosphere.  However,
compression-induced electron captures, neutron emissions, and
pycnonuclear reactions in the deep crust \cite{haensel90} will
maintain the core at a temperature $\approx 10^8
\langle\dot{M}/10^{-10}\,M_\odot{\rm\,yr^{-1}}\rangle^{0.4}\rm\,K$
\cite{bildsten97}.  A core at this temperature will make the NS
incandescent even after accretion halts (Brown, Bildsten \& Rutledge
\citenp*{brown98}), at a level of \ee{32}$-$\ee{33} \cgslum .  This
luminosity is {\em unavoidable\/}, unless neutrino emission is much
stronger than the standard modified Urca (such as may occur from a
pion condensate in the NS core; cf. \citenp{zdunik92}).

A resolution of the conflict between this expected thermal
emission and both the inferred small emitting area and the variability
of the quiescent luminosity is clearly needed.  In this paper, we
demonstrate that both observational objections can be alleviated by
consideration of two points.

First, the emitted spectrum from a non-accreting, low-magnetic field
NS atmosphere is not that of a blackbody.  The strong surface gravity
quickly ($\sim 10\second$) stratifies the atmosphere (cf.
\citenp{alcock80,romani87}).  For accretion rates $\lesssim 2\times
10^{-13}M_\odot {\rm\,yr^{-1}}$ (corresponding to an accretion
luminosity $\lesssim 2\times 10^{33}$ \cgslum), gravity removes metals
faster than the accretion flow supplies them (Bildsten, Salpeter \&
Wasserman \citenp*{bildsten92}).  As a result, the photosphere is
nearly pure hydrogen.  At temperatures $<0.5\keV$, the free-free
absorption, which is strongly frequency ($\nu$) dependent
(approximately $\propto \nu^{-3}$), dominates the opacity.  Because of
this frequency dependence, higher energy photons escape from greater
depths, where $T>\Teff$ \cite{pavlov78,romani87,zampieri95}.  Spectral
fits of the Wien tail with BB curves then overestimate $T_{\rm eff}$
and underestimate the emitting area, by as much as orders of magnitude
\cite{rajagopal96,zavlin96}.  Application of these models to the
isolated neutron star in SNR~PKS~1209-52 resulted in a source distance
consistent with that measured through other means (assuming a 10~km NS
radius), a lower surface temperature, and an X-ray measured column
density which was consistent with that measured from the extended SNR
(while the column density measured with an assumed BB spectrum was not
consistent; \citenp{zavlin98}).  

The second overlooked point is that there are large systematic
uncertainties in the equivalent hydrogen column density (\nh).
Successive X-ray observations of \x1608\ in outburst, separated by
$\sim$1.5 years, produced different galactic absorptions toward the
transient (\nh=(1.0\ppm0.1)\tee{22}\percm and
(1.5\ppm0.1)\tee{22}\percm ).  The change was attributed to outflows
from the transient X-ray source itself \cite{penninx89}.  If so, the
changes in the column depth are sufficient ($\delta\nhtt\sim 0.5$,
where $\nhtt\equiv\nh/\tento{22}\percm$) to account for the reported
variabilities in the quiescent luminosity of \cenx4\
(VP87; \citenp{verbunt94}).  For example, a change from $\nhtt=0.3$ to
$\nhtt=0.8$ will alter the observed flux from a BB spectrum of
temperature 0.3\keV\ by a factor of 2.5 in the 0.1-1.5\keV\ passband.

To summarize, the NS core, heated to a temperature of $T\sim10^8$ K,
emits a thermal spectrum which, if mis-interpreted as due to a BB
instead of a pure H-atmosphere spectrum, implies an emitting area
smaller than the area which actually produced the spectrum. A
reanalysis of the X-ray spectra obtained from quiescent NSs, using the
H atmosphere models of Zavlin \etal\ \cite*{zavlin96} is thus
warranted, to investigate that the spectral fits produce the expected
larger emitting areas, to check for the consistency of the size of the
emitting areas in the observed spectrum with a 10~km radius NS, and to
consider possible systematic uncertainties, such as the equivalent
hydrogen column density.

In Sec.~\ref{sec:data}, we describe the methodology used to re-analyze
the archival data.  We then re-fit the data for \cenx4\
(Sec.~\ref{sec:cenx4}), \aql\ (Sec.~\ref{sec:aqlx1}), and \x1608\
(Sec.~\ref{sec:1608}),  and show that the re-fitted emitting areas are
consistent with a NS surface area.  In Sec.~\ref{sec:conclude}, we
discuss these results and present conclusions.

\section{Data and Analysis}\label{sec:data}

Brief descriptions of the analyzed observations are in
Table~\ref{tab:datasets}.  Data were obtained from the public archive
at HEASARC/GSFC ({\tt http://heasarc.gsfc.nasa.gov/}).  We used two
separate observations of \aql\ while it was in quiescence using the
\rosat/PSPC, which have been previously analyzed \cite{verbunt94},
plus two observations of \aql\ using \asca : one while it was in
quiescence and the other in outburst (which we analyzed to obtain a
measurement of the column density toward \aql ).  We also re-analyze
\asca\ observations of \x1608\ and \cenx4\ in quiescence
\cite{asai96b}.  We did not attempt to re-analyze quiescent
observations of MXB~1730-335 (the Rapid Burster), as contamination
from a nearby object was found in the original investigation to make
spectral analysis infeasible \cite{asai96a}.

All X-ray spectra were fit using the XSPEC X-ray spectral analysis
package \cite{xspec}, using the standard BB (the model known as
``bbodyrad'') or a tabulated H-atmosphere model \cite{zavlin96}; the
galactic hydrogen column density (model known as ``wabs''), and, when
needed, the power-law (``powerlaw'').  The tabulated H-atmosphere
model used was for a neutron star with surface gravity $2.4\times
10^{14}\cm\second^{-2}$ (appropriate for a NS of gravitational mass
$1.4M_\odot$ and radius $10\km$), and a pure hydrogen atmosphere, with
the effective unredshifted temperature ($kT$) and apparent emission
area radius left as free parameters.  The spectral resolution and
signal-to-noise of the present data are insufficient to constrain the
additional parameters of NS mass, radius, and material metallicity.  By
fitting to this model, we are searching for consistency of the data
with a 10~km NS emission area radius.  Note especially that the $kT$
for the H-atmosphere model is the ({\it un}-redshifted) surface
temperature of the NS while the $kT_{\rm BB}$ is the temperature of
the NS as observed at an infinite distance (\ie\ the effect of
redshifting is not removed).  For a 10~km, 1.4\msun~NS, the redshift
factor to correct the BB temperature is 0.76.

\label{sec:absorption}

A perhaps dominant systematic uncertainty in spectral fits of this
type is the uncertainty in the equivalent hydrogen column density.
The \nh\ is strongly covariant with other parameters in an assumed BB
spectral model of $kT_{\rm BB}\sim0.3$ keV, and becomes a particular
problem in low S/N data with few spectral channels below 1.0 keV
where, if \nh\ is left as a free variable during spectral analysis, it
can produce relative uncertainties of order unity in other parameters.
The effect on spectra of NSs in quiescence with BB temperatures of
$\sim$ 0.1--0.3 keV is that the \nh\ is strongly covariant with the
object size, with lesser effect on the object temperature.  Thus, it
is not unusual for observers to hold this parameter fixed at a \nh\
value found from earlier observations (either when the X-ray source
was bright, or taking the value from optical measurements of $E(B-V)$
and using an average conversion factor found statistically from
measurements of the $E(B-V)$ and \nh\ toward other objects;
\citenp{predehl95,gorenstein75}), which produces smaller error bars on
other parameters. This is done based on the assumption that the column
density is due largely to material distributed between the observer
and the object, and is not affected by gas that the object may eject
into the surrounding environment -- or, if it is, that the amount
which is ejected does not vary over time.

As we note in Sec.~\ref{sec:intro}, X-ray observations during
successive outbursts of \x1608\ have measured values of \nh\ which
differed by \nhtt$\sim$0.5.  Changes of this same magnitude in \nh\
toward \aql\ during an outburst over short time-scales have also been
reported \cite{czerny87}; differences were noted between the measured
column density in the tail of type~I X-ray bursts when compared with
the column density in the 2000 sec prior (\nhtt=0.4 vs. 1.0), and
observed decay in this parameter in $\sim$ 1000~s following the type~I
burst.

In addition to a possible time variability in \nh, there are
systematic uncertainties in calibration of the X-ray absorption
(mostly from photoelectric absorption in metals) with optical
reddening (from dust).  The X-ray absorption can, in principle, be
estimated from reddening of the binary companion or a nearby star
because the equivalent hydrogen column density \nh\ strongly
correlates with the optical extinction $A_{\rm V}$. For instance,
Predehl \& Schmitt \cite*{predehl95} obtained \nhtt$=0.179 A_V$ (with
a formal uncertainty of 0.003) from \rosat\ observations of X-ray
halos around persistent sources.  However, the value of the conversion
factor depends on the average dust-to-gas ratio along the line of
sight, which may be different for different directions and distances;
this is demonstrated by considerable scatter of individual points in
the \nh--$A_{\rm V}$ diagrams \cite{savage79}.  In addition, the value
of the conversion factor obtained by Predehl \& Schmitt
\cite*{predehl95} is different from that obtained by Gorenstein
\cite*[\nhtt$=0.222 A_{\rm V}$]{gorenstein75}, which was attributed 
to systematic effects in both studies.  Thus, systematic uncertainties
in this conversion factor are at least $\sim$25\%.

In the specific case of \aql, a measurement of $E(B-V)=0.37$ mag
\cite{thorstensen78} (implying \nhtt=0.205\ppm 0.003, using the formal
\nh-$A_{\rm V}$ uncertainty) is based on the optical reddening of the
quiescent K0 counterpart.  However, it was noted by the authors that
this measurement was discrepant with that of a normal, nearby ($1\farcm4$)
B-star, for which optical reddening was measured to be $E(B-V)=0.73$
mag, implying \nhtt=0.404\ppm 0.006.  This discrepancy is greater
than the formal uncertainty and the $\sim$ 25\% systematic uncertainty
in the $A_{\rm V}$-\nh\ conversion.  As we show in
Sec.~\ref{sec:aqlx1}, the higher column density is consistent with
that which we measure with an assumed X-ray spectrum during an X-ray
outburst. The difference in \nh\ from these measurements translates
into a factor of two in the fitted emission area radius; while
important, this uncertainty is smaller than the difference between a
BB and a H atmosphere.

One must also keep in mind that the magnitude of the X-ray measured
column density depends on the assumed spectrum as well; if the assumed
spectrum is not the correct intrinsic spectrum, then the X-ray
measured column density can be different from its true value. 

If the column density can be reliably measured during an X-ray
observation, then the spectral parameters which are covariant with the
column depth can also be reliably measured. However, for the presently
investigated observations with the \rosat/PSPC and \asca, the
signal-to-noise is not high enough to measure the column density to
within a factor of two, which results in a systematic uncertainty in
the derived emitting radius of a factor of 2 or more.  To investigate
how this uncertainty affects the interpretation of our spectral
results, we adopt the practice of freezing \nh\ at a range of values,
some historically measured, others higher than these by \nhtt=0.5.
While this complicates interpretation of the best-fit spectral models,
it does not diminish our ability to investigate the systematic
differences in X-ray emitting areas between BB and H-atmosphere
models.

\section{Sources}

The results of our spectral fits are presented in Table
~\ref{tab:results}.  This table contains: (1) the dataset number
(cf. Table~\ref{tab:datasets}); (2) the assumed (or fit) \nh ; (3) the
best-fit spectral parameters for the H-atmosphere model, including the
{\it un}-redshifted effective NS surface temperature ($kT$) and
apparent emission area radius ($r_e$), as well as the reduced
$\chi^2_\nu$ for that model; (4) the best fit spectral parameters for
the BB model, including the redshifted NS surface temperature
($kT_{\rm BB}$), and emission area radius $r_e$, and the reduced
$\chi^2_\nu$ for that model; (5) when used, the best fit photon
power-law spectral parameters, including the photon power-law spectral
slope ($\alpha$) and model normalization.

 For compactness, we sometimes list both the best-fit H-atmosphere
model on the same line as the best fit BB model, when we used the same
value of \nh\ for these.  We do not list the two separate models on
the same line when we also add to the fit a power-law spectral model,
to clarify that the parameters of the best-fit power-law spectral
model are different depending on whether one uses the H-atmosphere
model or the BB model.  Finally, when we tie together all but one of
the model parameters of three different data-sets (in the case of
\aql) we use the caret (``) on subsequent lines to indicate this,
while the parameter which is permitted to vary between the three
data-sets is listed distinctly on the line corresponding to that
data-set.

In general, for assumed values of \nh\ the H-atmosphere emission area
radii are systematically larger than those of the BB model, by factors
between 4--10.  While the exact radius implied depends on the assumed
\nh , those of the BB model cannot be made consistent with a
$\sim$10~km~NS emission radius for the range of \nh\ we assume here
based on past observations and allowing for some variability, while
those of the H-atmosphere model can be made consistent with a 10~km NS
emission radius.

\subsection{Cen X-4}\label{sec:cenx4}

A previous optical reddening observation in \cenx4\ ($E(B-V)$=0.1 mag;
\citenp{blair84}), implies a column density of \nhtt=0.055, consistent
with the best published X-ray constraint on this value from these observations
(\nhtt$<$0.03--0.2, depending on assumed spectrum; \citenp{asai96b}).

We used the standard data products spectra for the SIS0, SIS1, GIS2,
and GIS3 detectors. For the GIS2+3 data we subtracted background taken
in four, $5'$ radius circular regions surrounding the source, with no
region overlapping any area within $5\farcm$ of the source, using the
screened events.  For the SIS0+1 data, we subtracted background from
two rectangular regions, one about $1\farcm9\times12'$, the other
$2\farcm5\times5\farcm2$, whose edges were $2\farcm5$ away from the object at
their closest, and at least 10 pixels from the edge of the detector,
using the screened events.  We ignored energy channels below 1.0 keV
in the GIS2+3 detectors, to conservatively avoid energy regions of
calibration uncertainties.

As found previously by Asai \etal \cite*{asai96b}, a power-law is
required to account for emission above 5 keV (without it, the best fit
$\chi^2$ is 250 for 130 degrees of freedom).  

We successively hold the \nh\ constant at the optical reddening value
\nhtt=0.055, and at the 2$\sigma$ upper limit (\nhtt=0.2;
\citenp{asai96b}).  The best fit BB spectrum is consistent with that
found previously with the same data.  For the BB model, the implied
apparent radius is 1.2\ud{0.15}{0.2} km for \nhtt=0.055, and is
2.9\ud{0.7}{0.5} km for \nhtt=0.2.  For the H-atmosphere model, the
implied apparent radius is 7.8\ud{1.9}{1.7} km when \nhtt=0.055, and
is 32\ud{13}{8} km when \nhtt=0.2.

For the best-fit H atmosphere model with \nhtt=0.055, the unabsorbed
(0.5--10.0 \keV) luminosity is (1.6\ppm0.6)\tee{32} $(d/1.2 \kpc)^2$
\cgslum\ , for the thermal component only.  

\subsection{\aql}\label{sec:aqlx1}

For the \rosat observations (\#2 and \#3;
cf.~Table~\ref{tab:datasets}), we extracted the source spectrum from
within a circle $30'$ in radius, centered on the object.  The
background was taken from a nearby $200''$ circular region in the
inner part of the PSPC detector.  For observation \#2, spectral fits
of a BB with galactic absorption were consistent with those found
previously with the same data \cite{verbunt94}. We find a source
emission region of apparent radius of 0.62\ud{0.12}{0.10}~km (for
\nhtt=0.2; 90\%), increasing to 1.3\ppm~0.3~km for \nhtt=0.4.  Using
the H-atmosphere model, we find the emission region has an apparent
radius of 2.4\ud{0.9}{0.6}~km (\nhtt=0.2), which increases to
8.7\ud{4.1}{2.7}~km for the higher assumed absorption (\nhtt=0.4).
These results are consistent with those of observation \#3.  For an
assumed spectrum of a black-body with \nhtt=0.4, we find a flux of
2.9\tee{-13} \cgsflux (0.5-2.0 keV); for observation \#3, the same
assumed spectrum produces a flux of 5.5\tee{-13} \cgsflux.
 
For the \asca\ quiescent observation (\#4) of \aql , we used the
standard data products source energy spectrum from the SIS0+1 and
GIS2+3 detectors.  For the GIS background, we used three circular
areas $5'$ in radius, each centered approximately at an equal distance
from the GIS detector center as the source, and which do not overlap
any area to within $5'$ of the source, using the screened events.  We
excluded all energy channels below 1.0 keV for the GIS data from the
fit. For the SIS background, we used 3 rectangular areas, each at
least $2.5'$ away from the source center, and 10 pixels from the edge
of the detector, using the screened events (we examined a background
spectrum using only detector area at least $3'$ from the source, and
found no significant difference between the two background spectra).
The measured BB spectrum produces a smaller area than that found from
the \rosat/PSPC observations (0.57\ud{0.10}{0.08} km for \nhtt=0.40)
and a higher temperature (0.32 \ppm 0.02 keV). As with the \rosat/PSPC
observations, the larger assumed column density results in a larger
emission area, which is marginally consistent with the \rosat/PSPC
measurement.  The H-atmosphere model produces a larger emission area
radius than the BB model (2.6 km vs. 0.57 km, for an assumed distance
of 2 \kpc).  However, to produce a radius consistent with $\sim$ 10
(d/2 \kpc) km, the column density must be \nhtt$\sim$ 0.8.

\aql\ was observed during a bright phase with \asca\ (12.76 \ppm 0.02
c/s in GIS2; Observation \#5).  For this observation, we used the
standard products X-ray spectrum for GIS2+3 data, using only medium-
and high-bitrate PH data, and neglected the background. (The SIS0+1
data were telemetry saturated).  We excluded energy channels below 1.0
keV from the fits. The dead-time of the GIS2 was $\sim$12\% and the
GIS3 was $\sim$14\% due to telemetry saturation, which affects the
normalization of the models (which have been approximately corrected),
but not the parameters ($kT$, \nh).  We fit the average X-ray spectrum
measured by the GIS2+3 detectors during this period with a BB plus
power-law model with an equivalent hydrogen column density, which
produced a column density of \nhtt = 0.425 \ppm 0.02 (90\%).  This
value is comparable to those found for this object previously
(\nhtt=0.53\ppm0.01, and 0.36\ppm0.02; \citenp{christian97}).  The
best fit model corresponds to an average flux of $1\times 10^{-9}$
ergs \percm \persec (0.5--10.0 keV), and an unabsorbed average
luminosity (0.5--10 keV) of $6.9\times 10^{35}$ ergs \persec (d/2
kpc)$^{2}$.

We investigated the consistency of all three quiescent data-sets
having been produced by the same spectral model.  We fit these
simultaneously, using galactic absorption and a blackbody model.  If
we fix all three parameters simultaneously, the best fit model is
formally unacceptable, with a probability ($p$) of producing a dataset
this discrepant with the intrinsic spectrum of $p$=\ee{-6}, with a
\chisqr\ of 167 for 94 degrees of freedom. If we allow the column
densities of the three observations to vary independently of one
another while tying together the temperature and emitting area, we
obtain a statistically acceptable fit ($p$=0.93) and similar column
densities between observations 2 and 4, but a different one (by
\nhtt=0.21\ppm 0.11; 90\% confidence) for observation 3.  However, if
we instead allow the BB temperatures to vary independently of one
another while tying together the column density and emitting area, we
also obtain a statistically acceptable fit ($p$=0.37), and the
temperatures are consistent with one another at the 90\% confidence
level.  Similarly, if we allow the emitting areas to vary while tying
the column densities and temperatures together, the emitting area
radii are consistent at the 90\% confidence level.  Thus, there is
some variation in the intrinsic spectrum between these three
observations, which could be caused by significant (at the$>$90\%
level) changes in the column density, or by small changes in the
either the temperature or the emitting area, or by a combination of
changes in all three.  Performing the same analysis using a H
atmosphere X-ray spectrum with galactic absorption produces similar
results.

We performed a combined fit of all three datasets of \aql\ in
quiescence using galactic absorption and the H atmosphere model. When
all three datasets are fit with the same spectral model, the spectra
are again found to be inconsistent with being identical
($p$=7\tee{-6}; $\chi^2$=150 with 82 dof).  When \nh\ is permitted to
vary between the three datasets, but the intrinsic spectrum is held to
be the same for all three, we find an acceptable fit, with the
apparent radius $r_e$=3.5\ud{2.0}{1.2} (d/2 \kpc) km (90\%).  This
value is consistent with that expected from a 10~km NS, but only if
the distance to \aql\ is toward the higher part of the 1.7-4.0 \kpc\
range found by optical spectral-typing of the secondary
\cite{thorstensen78}.  Thorstensen \etal\ also argued that the
discrepant reddening they observe between \aql\ and a nearby B~star
would be resolved if the companion in \aql\ were bluer than the median
spectral type they asssumed, thus more luminous and more distant,
within their 1.7-4.0 \kpc\ estimate.  If the companion to \aql\ is at
4.0\kpc , the apparent radius is then 7.0\ud{4.0}{2.4} km for the H
atmosphere model.  We interpret this result as favoring a 4\kpc\
distance to \aql.

In the best combined-fit for the three datasets using a H atmosphere
model with \nh\ permitted to vary but with the intrinsic spectra of
the three models constrained as identical, the 0.5--10.0 keV unabsorbed
luminosity was (5.1\ppm2.3)\tee{32} $(d/ 2\kpc)^{2}$ \cgslum.

\subsection{\x1608}\label{sec:1608}

We used the standard data products spectra for the SIS0+1 and GIS2+3
detectors.  For the GIS2+3 background we used an annulus centered on
the source of inner radius $3'$ and outer radius $12'$, using the
screened events.  For the SIS0+1 data, it was found by Asai \etal\
\cite*{asai96b} that galactic ridge emission contributed significantly
to the background, which required an off-source observation to
estimate; we used the identical background spectrum for the present
spectral analysis (kindly provided by K. Asai).  For the GIS2+3 data,
we ignored all energy channels below 1.0 keV. For the SIS0+1 data, we
ignored all energy channels below 0.5 keV.  Becayse of the lack of
counts, we ignored all energy channels above 5 keV \cite{asai96b}.

The column density toward \x1608\ has been measured to be variable
\cite{penninx89}, between \nhtt=1.0 and 1.5 (\ppm0.1).  The measured
optical reddening $E(B-V)$=1.5 mag \cite{grindlay78} implies a column
density of \nhtt=0.8.  We assumed alternately \nhtt=0.8 and 1.5.

For the low value of \nhtt=0.8 , the BB model produces an emission area
radius of 1.7\ud{0.5}{0.3}~km, while the H-atmosphere model produces
9.4\ud{4.5}{2.7}~km.  For the high value of \nhtt=1.5, the BB-model radius
is 4.6\ud{1.6}{1.1}~km, while the H-atmosphere model radius is
50\ud{32}{20}~km.

For the best-fit H atmosphere model, with \nhtt=0.8, the unabsorbed
(0.5--10.0 \keV) luminosity is (8.3\ppm4.2)\tee{32} $(d/3.6 \kpc)^2$
\cgslum; and for \nhtt=1.5, it is (7.3\ppm4.7)\tee{32} $d/3.6
\kpc)^2$. 

\section{Discussion and Conclusions}\label{sec:conclude}

Using realistic hydrogen atmosphere models and accounting for
systematic uncertainties in the absorption, we have reanalyzed the
quiescent spectra of the transient NSs \aql, \cenx4, and \x1608.  We
find that the emission areas are always larger (by factors of 16-100)
than implied by spectral fits that use a blackbody spectrum. These
emitting areas are, within the large systematic uncertainty due to \nh
, consistent with the surface area of a NS, and suggest that a
substantial fraction of the quiescent luminosity is thermal emission
from the NS surface. Even allowing for large deviations in the
hydrogen column density cannot make the emission area inferred from BB
fits commensurate with that of a NS.

If the thermal component of the quiescent spectra originates from the
NS surface, then two likely causes are incandesence from the hot core
\cite{brown98} or accretion at low rates onto much of the NS surface
\cite{zampieri95}.  The latter interpretation requires a small
magnetic field or a slow rotation rate to avoid a propeller effect
\cite{ill75}. For \aql, the rotation period must be
$>0.6\,(B/10^9\gauss)^{6/7}\second$ \cite{verbunt94} to allow
accretion in quiescence. Interpreting the 549~Hz oscillation seen
during a type~I burst from \aql\ \cite{zhang98a} as the spin frequency
then implies that $B<10^6\gauss$.  Alternatively, if accretion in
spite of the propeller should occur, then magnetic funneling of the
accretion flow onto the polar caps may produce variability in the
thermal component at the NS spin frequency, as with accretion-powered
pulsars \cite{zhang98b}.  No such variability has been reported.

For \cenx4\ and \aql , an additional emission mechanism is required to
explain the hard power-law tail.  This hard emission might emanate
from the magnetopause, or from the interaction of a pulsar wind with
ambient material, such as has been seen in the pulsar/Be star system
PSR~1259--63 \cite{tavani97}.  There are presently no proposed models
in which these mechanisms produce a thermal spectrum in addition to
the hard power-law tail, in particular one with an emitting area
comparable to that of a NS; thus the power-law spectral component may
exist -- and its properties may vary -- independently of the thermal
component.

The large systematic uncertainty in the hydrogen column density
dominates the results we present here.  Unfortunately, the difficulty
in determining \nh\ simultaneously with other spectral parameters in
these low S/N data cannot be overcome by assuming historically
measured column densities, because these are systematically uncertain
by $\nhtt\sim 0.5$, which is sufficient to change the implied radius
by a factor of two, as are the distance measurements.  Improved
modeling of the X-ray binary optical spectra can perhaps correct the
$\nh\mbox{--}A_{\rm V}$ relation, but the uncertainty from time
variability of the column will still remain.  The only remedy is high
signal-to-noise X-ray spectral data in a passband that covers both the
energy range affected (0.5--2.0\keV) and the energy range unaffected
($>2\keV$) by the absorption depths common to these sources
($\nhtt\approxlt 1$). Both \axaf\ and \xmm\ should be capable of these
kinds of observations.

\acknowledgements

This research was supported by NASA via grant NAGW-4517 and through a
Hellman Family Faculty Fund Award (UC-Berkeley) to LB.  EFB is
supported by NASA GSRP Graduate Fellowship under grant NGT-51662. GGP
acknowledges support from NASA grants NAG5-6907 and NAG5-7017.  This
research has made use of data obtained through the High Energy
Astrophysics Science Archive Research Center Online Service, provided
by the NASA/Goddard Space Flight Center.  We are grateful to K. Asai
for providing the background files for the analysis of \x1608\ with
the \asca/SIS instruments.

\newpage

\begin{deluxetable}{lcccc}
\tablewidth{40pc}
\tablecaption{X-ray Observations\label{tab:datasets}
 }
\tablehead{
Number	& Satellite/Instrument&	Obs Start Time		& Live Time 	& Avg Countrate\tablenotemark{a} \nl
	&		      & (UT)			& (ksec)	& (c/s) \nl}
\startdata
\multicolumn{5}{c}{ \cenx4} \nl
1	&\asca/SIS+GIS		&27/02/94 04:11		& 28.0	& 0.017\ppm 0.001 (GIS2)\nl \tableline
\multicolumn{5}{c}{\aql} \nl
2	&\rosat/PSPC		&15/10/92 13:19		& 14.4 	& 0.029 \ppm 0.002 (PSPC)	\nl
3	&\rosat/PSPC		&24/03/93 04:41		& 12.5 	& 0.055 \ppm 0.002 (PSPC)	\nl
4	&\asca/SIS+GIS		&21/10/96 11:00		& 37.6	& 0.012 \ppm 0.001 (GIS2)	\nl
5	&\asca/GIS		&30/04/94 15:25		& 30.2	& 12.76\ppm 0.02 (GIS2) \nl \tableline
\multicolumn{5}{c}{ \x1608} \nl 
6	&\asca/SIS+GIS		& 12/08/93 05:49	& 34.3	& 0.013\ppm0.001 (GIS2) \nl \tableline	
\tablerefs{(1)  \citenp{asai96b}; (2) \citenp{verbunt94}; (3) \citenp{verbunt94}; (4) and (5), the present work;  (6) \citenp{asai96b}}
\tablenotetext{a}{ Count-rates are background subtracted. For ASCA,
count-rates are for GIS2 detector (0.8-10 keV), and for the \rosat/PSPC,
energy range is (0.4-2.4 keV)}
\enddata
\end{deluxetable}

\begin{table}[htb]
\begin{tiny}
\begin{center}
\tablewidth{40pt}
\caption{X-ray Spectral Parameters of \aql , \cenx4, \& \x1608\label{tab:results}}
\begin{tabular}{cccccccccccc} \tableline \tableline
Dataset        &  \nh             &\multicolumn{3}{c}{H Atmosphere}  & & \multicolumn{3}{c}{Ideal Blackbody}	&	$\alpha$\tablenotemark{b}	& Norm\tablenotemark{c}   \\ \cline{3-5} \cline{7-9}
Number 		& 		&$kT$ & $r_e$  &$\chi^2$/dof& & $kT_{\rm BB}$& r  &$\chi^2$/dof  			& 			& 		\\
		&{(\ee{22} \percm)}& {(keV)} 	& {(km ($d/d_0$))\tablenotemark{a}}&	& & {(keV)} 	& {(km ($d/d_0$))\tablenotemark{a}}&      & \\ \tableline \tableline
\multicolumn{11}{c}{ \cenx4 } \\						 
1 		&(0.055)		&...		&...		&...	&&0.180\ppm0.0015&1.2\ud{0.15}{0.2}&119/127&1.5\ppm0.35		&1.0\ud{0.5}{0.35}\\ 
		&(0.2)			&...		&...		&...	&&0.144\ppm0.012&2.9\ud{0.7}{0.5}&126/127&1.85\ud{0.4}{0.3}	&1.7\ud{0.8}{0.5}\\
		&(0.055)		&0.10\ppm0.012	&7.8\ud{1.9}{1.7}&124/129&& ...		&...		&...	&1.07\ud{0.3}{0.15}	&0.55\ud{0.45}{0.25}\\
		&(0.2)			&0.063\ud{0.07}{0.08}&32\ud{13}{8}&118/129&& ...		&...		&...	&1.4\ud{0.45}{0.35}	&1.0\ud{0.6}{0.4} \\ \tableline \tableline
\multicolumn{11}{c}{ \aql } \\
2	  	&(0.2)			&0.19\ud{0.02}{0.03}&2.4 \ud{0.9}{0.6}&25/18&&0.29\ppm 0.02	&0.62\ud{0.12}{0.10}	&21/18	&...		&...	\\
		&(0.4)			&0.12 \ppm 0.02	&8.7\ud{4.1}{2.7} &18/18&&0.22\ppm 0.02	&1.3 \ppm 0.3			&17/18	&...		&...	\\
3	  	&(0.2)			&0.205 \ppm 0.02&2.7\ud{0.7}{0.6}&30/18	&&0.295\ppm 0.02&0.75\ud{0.12}{0.10}		&26/18	&...		&...	\\
		&(0.4)			&0.13\ppm 0.015	&9.3\ud{3.1}{2.3}&19/18	&&0.232\ppm 0.015 &1.6 \ud{0.29}{0.24}		&17/18	&...		&...	\\
4		& (0.2)			&0.25\ppm 0.025	&1.5\ud{0.4}{0.3}&39/56&&0.36\ppm 0.02	& 0.40\ud{0.06}{0.05}	&40/56	&...		&...	\\
		& (0.4)			&0.20\ppm 0.025 & 2.6\ud{0.7}{0.6}&38/56&&0.32\ppm 0.02	& 0.57\ud{0.10}{0.09}	&43/56 &...		&...	\\
		& (0.8)			&0.14\ppm 0.023	& 7.5\ud{3.0}{2.2}&49/56 &&0.27\ppm 0.02	& 1.10\ud{0.25}{0.19}	&59/56 &...		&...	\\
2, 3, \& 4 			& 0.15\ppm 0.06	&...			&...		&...	&&0.346\ppm 0.025	&0.43\ud{0.10}{0.07}	&161/82	&...		&...	\\ 		\tableline
2, 3, \& 4	& 0.20\ud{0.08}{0.06}		& 0.22\ppm0.03	&2.0\ud{0.8}{0.6}	&150/82	&&...			&...			&...	&...		&...	\\ 		\tableline
\multicolumn{1}{r}{2}		& 0.44\ud{0.12}{0.10}  	&0.18\ppm0.03	&3.5\ud{2.0}{1.2}&87/80 &&...			&...			&...	&...		&...	\\ 
\multicolumn{1}{r}{(\nh vary) 3}& 0.22\ppm0.08		& ``		&	``	& `` 	&&...			&...			&...	&...		&...	\\ 
\multicolumn{1}{r}{ 4}		& 0.44\ud{0.12}{0.11}	& ``		&	``	& ``	&&...			&...			&...	&...		&...	\\ 
\multicolumn{1}{r}{2}		& 0.35\ud{0.12}{0.08}  	&...		&...		&...	&&0.311\ppm 0.025	&0.61\ud{0.19}{0.12}	&101/80	&...		&...	\\
\multicolumn{1}{r}{(\nh vary) 3}& 0.14\ud{0.07}{0.06}	&...		&...		&...	&& ``			&	``		& ``	&...		&...	\\
\multicolumn{1}{r}{ 4}		& 0.35\ud{0.12}{0.09}	&...		&...		&...	&& ``			&	``		& ``	&...		&...	\\ \tableline
\multicolumn{1}{r}{2}           & 0.16\ppm 0.06		&...		&...		&...	&&0.31\ud{0.03}{0.025}	&0.48\ud{0.13}{0.09}	&90/80	&...		&...	\\
\multicolumn{1}{r}{($kT_{\rm BB}$ vary)   3}& ``	&...		&...		&...	&&0.36\ud{0.04}{0.03}	& ``			& ``	& ...		& ...	\\
\multicolumn{1}{r}{ 4}          & `` 			&...		&...		&...	&&0.33\ud{0.03}{0.027}	& ``			& ``	& ...		& ...	\\ \tableline
\multicolumn{1}{r}{2}		& 0.17\ud{0.07}{0.06}	&...		&...		&...	&&0.336\ppm 0.028	&0.42\ud{0.11}{0.08}	&83/80	&...		&...	\\
\multicolumn{1}{r}{(r vary) 3}	& `` 			&...		&...		&...	&& ``			&0.56\ud{0.15}{0.10}	& ``	& ...		& ...	\\
\multicolumn{1}{r}{ 4}          & `` 			&...		&...		&...	&& ``			&0.46\ud{0.12}{0.09}	& ``	& ...		& ...	\\ \tableline
5 		&0.425\ppm 0.02		&	...	&	...	&	&&0.94\ud{0.04}{0.03}& 0.85\ud{0.08}{0.09}&1954/1461&1.635\ppm0.035& 1900\ppm90	\\ \tableline \tableline
\multicolumn{11}{c}{ \x1608 } \\						 
6 			&(0.8)		&0.17\ppm0.03   &9.4\ud{4.5}{2.7}&70/61 &&0.30\ud{0.2}{0.3}&1.7\ud{0.5}{0.3}&70/61&...			&...	\\ 
			&(1.5)		&0.105\ud{0.02}{0.016}&50\ud{36}{20}&74/61&&0.235\ppm0.020&4.6\ud{1.6}{1.1}&39/61	&...			&...	\\ \tableline \tableline
\tablenotetext{}{Errors are 90\% confidence; Values listed in
parenthesis were fixed during the fits}\\
\tablenotetext{a}{ Assumed Distances: \aql , $d_0$ =2kpc \cite{czerny87}; \x1608 , $d_0$=3.6 kpc \cite{nakamura89}; \cenx4 $d_0$= 1.2 kpc
\cite{chev89}}
\tablenotetext{b}{ Power-law Photon Slope}
\tablenotetext{c}{ Normalization in \ee{-4} photons~\percm~\persec~keV$^{-1}$ at 1 keV.  }
\end{tabular}
\end{center}
\end{tiny}
\end{table}

\end{document}